\documentclass[12pt,preprint]{aastex}

\slugcomment{To be appeared in AJ}

\shorttitle{Spectropolarimetry of Nova V475 Sct}
\shortauthors{Kawabata et al.}

\begin{document}

\title{Low- and Medium-Dispersion Spectropolarimetry of Nova V475 Sct
 (Nova Scuti 2003): Discovery of an Asymmetric High-Velocity Wind in
 a Moderately Fast Nova\footnotemark[1]}

\footnotetext[1]{Based on data obtained at the Subaru Telescope
and the Okayama Astrophysical Observatory, which are operated
by the National Astronomical Observatory of Japan (NAOJ)}

\author{
Koji S. Kawabata\altaffilmark{2},
Youichi Ohyama\altaffilmark{3,4},
Noboru Ebizuka\altaffilmark{5},
Tadafumi Takata\altaffilmark{3,6},
Michitoshi Yoshida\altaffilmark{7},
Mizuki Isogai\altaffilmark{8},
Yuji Norimoto\altaffilmark{7},
Akira Okazaki\altaffilmark{9}, and
Masashi S. Saitou\altaffilmark{9}
}

\altaffiltext{2}{Hiroshima Astrophysical Science Center,
Hiroshima University, 1-3-1 Kagamiyama, Higashi-Hiroshima,
Hiroshima 739-8526, Japan; kawabtkj@hiroshima-u.ac.jp.}

\altaffiltext{3}{Subaru Telescope, NAOJ,
650 North A'ohoku Place, Hilo, HI 96720, USA;
ohyama@naoj.org, takata@naoj.org.}

\altaffiltext{4}{Department of Infrared Astrophysics, ISAS,
Japan Aerospace Exploration Agency (JAXA),
3-1-1 Yoshinodai, Sagamihara, Kanagawa 229-8510, Japan.}

\altaffiltext{5}{Integrated V-CAD System Research Program,
RIKEN (The Institute of Physical and Chemical 
Research), Wako, Saitama 351-0198, Japan; ebizuka@riken.jp.}

\altaffiltext{6}{Astronomical Data Analysis Center, NAOJ,
2-21-1 Osawa, Mitaka, Tokyo 181-8588, Japan.}

\altaffiltext{7}{Okayama Astrophysical Observatory, NAOJ,
3037-5 Honjo, Kamogata-cho, Asakuchi-gun, Okayama 719-0232, Japan;
yoshida@oao.nao.ac.jp, agk00226@v1.yct.ne.jp}

\altaffiltext{8}{Kiso Observatory, Institute of Astronomy,
University of Tokyo, Mitake-mura, Kiso-gun, Nagano 397-0101, Japan;
iso@kiso.ioa.s.u-tokyo.ac.jp.}

\altaffiltext{9}{Department of Science Education, Gunma University,
Aramaki 4-2, Maebashi 371-8510, Japan; okazaki@edu.gunma-u.ac.jp,
saitou-m@ed.edu.gunma-u.ac.jp.}

\begin{abstract}
We present low-resolution ($R\sim 90$) and medium-resolution 
($R\sim 2500$) spectropolarimetry of Nova V475 Sct
with the HBS instrument, mounted on the 0.91-m telescope at the 
Okayama Astrophysical Observatory, and with FOCAS, mounted
on the 8.2-m Subaru telescope.
We estimated the interstellar polarization toward the nova
from the steady continuum polarization components and 
H$\alpha$ line emission components.
After subtracting the interstellar polarization component 
from the observations, we found that the H$\alpha$ emission
seen on 2003 October 7 was clearly polarized.
In the polarized flux spectrum, the H$\alpha$ emission had a
distinct red wing extending to $\sim +4900$ km s$^{-1}$ and
a shoulder around $+3500$ km s$^{-1}$,
showing a constant position angle of linear polarization
$\theta_{\rm *}\simeq 155\arcdeg\pm 15\arcdeg$.
This suggests that the nova had an asymmetric outflow with a
velocity of $v_{\rm wind}\simeq 3500$ km s$^{-1}$ or more,
which is six times higher than the expansion velocity
of the ionized shell at the same epoch.
Such a high-velocity component has not previously been reported for a
nova in the `moderately fast' speed class.
Our observations suggest the occurrence of violent mass-loss 
activity in the nova binary system even during the common-envelope phase.
The position angle of the polarization in the H$\alpha$ wing
is in good agreement with that of the continuum 
polarization found on 2003 September 26 ($p_{\rm *}\simeq 0.4$--$0.6$ \%),
which disappeared within the following 2 d.
The uniformity of the PA between the continuum polarization and 
the wing polarization on October 7 suggests that the axis of
the circumstellar asymmetry remained nearly constant
during the period of our observations.
\end{abstract}


\section{Introduction}

  It is generally accepted that the outburst of a classical nova
  is due to a thermonuclear runaway in a hydrogen-rich envelope
  accreted onto a white dwarf in a semi-detached binary system.
  The outburst ejects $\sim 10^{-4}$M$_{\sun}$ of matter from
  the white dwarf surface with velocities up to several thousand
  km s$^{-1}$ (see, e.g., \citealt{bod89}).
  The study of nova outburst has consequences for a number of areas, 
  including the study of nucleosynthesis in the accreted envelope,
  the processes of dust formation in the ejected matter, and
  the gas dynamics and chemical evolution of the interstellar medium.

  The shapes of the outburst ejecta provide an insight into the
  physical processes that occur during outbursts.
  Deep imaging and spectroscopic observations of old nova remnants
  have revealed that many of them have asymmetric structures,
  such as prolate shells including rings and blobs
  (e.g., \citealt{sla95,par95,gil00,kra02,har03}).
  In addition, polarimetry has revealed that the asymmetric
  geometry was already present at the outburst phase for many novae
  \citep{egg67,zel71,pii79,ber88,kik88,kuc90,bjo94,elk95,oka96,joh97,
  kaw01,eva02,wis03,des04}.
  However, despite the accumulation of observational facts,
  it is still not clear what gives rise to departures from
  spherical symmetry in nova ejecta.

  In a typical model of nova mass loss in the outburst phase, the
  ejecta are presumed to consist of an initially ejected main
  shell and a steadily blowing faster wind.
  In this model, the different expansion
  velocities of the shell and the wind should cause them to collide.
  Using such an interacting wind model, \citet{llo97} carried out
  hydrodynamic simulations of the ejecta and were able to explain
  some of the observed structures, such as polar blobs and equatorial bands.
  In their model, the secondary star experiences a frictional
  drag force inside the common envelope and transfers angular
  momentum to the ejecta (cf., \citealt{mac86,liv90}),
  causing a fast anisotropic flow that collides with the 
  outer spherical shell.
  However, the effects of a rapidly rotating white dwarf
  may also be important \citep{por98,sco00}.
  Centrifugal forces reduce the effective surface gravity
  of the white dwarf at the equator and the electron degeneracy
  at the base of the accreted envelope becomes stronger at the poles.
  The ejected material gains higher velocities in the polar
  regions than at the equator regions, which gives rise to
  a high-velocity bipolar outflow.
  More recently, a transient high-velocity emission component was
  found in Nova V1494 Aql during its transition phase
  \citep{iij03}, and \citet{ret04} suggested that a temporary
  increase in the accretion rate resulted in the formation of a jet
  at the inner boundary of a reformed accretion disk (cf., \citealt{sok04}).

  These studies assume or predict that the nova wind is highly
  asymmetric.
  If free electrons inside high-velocity winds scatter
  substantial amounts of the NLTE line emission from lower-velocity
  regions near the central star,
  redshifted, polarized light can be expected to contribute to the net flux,
  which will result in a broad red wing associated with an
  emission line in the polarized flux spectrum.
  Thus, spectropolarimetry with substantial wavelength
  resolution has the potential to provide us with unique information on the
  structure and kinematics of nova winds.

  Nova V475 Sct (= Nova Scuti 2003) was discovered by
  H. Nishimura on 2003 August 28.58 UT \citep{nak03}
  and reached its maximum brightness of $V_{\rm max}=8.43$
  on September 1.77 UT (JD = $2,452,884.27$; \citealt{cho05}, see
  also \citealt{str04} and \citealt{mor05}).
  In this paper, we consider this time as 0 d of the nova epoch.
  \citet{yam03} reported that the precise position is 
  $\alpha_{2000}=18^{\rm h}49^{\rm m}37\fs 600$ and
  $\delta_{2000}=-9^{\circ}33'50\farcs 85$ 
  ($l=24\fdg 20$, $b=-3\fdg 95$)
  and that no object is found within $2''$ on the 
  red 1984 DSS image or 1999 2MASS public images.
  The $V$-band lightcurve showed small irregular variations 
  in the early decline phase, followed by a deep decline 
  at $\sim 60-200$ d \citep{cho05}.
  The early decline rate of V475 Sct was $t_{2}\simeq 48$ d
  \footnote{\citet{mor05} derives $t_{2}=22$ d from AAVSO light curve
  data, which is considerably shorter than that derived by
  \citet{cho05}. In this paper, we adopt the latter because
  the maximum brightness $V_{\rm max}=8.0\pm 0.2$ in \citet{mor05}
  is based on the very short flare seen in the curve around
  September 1.8 UT and such a short variation would have been
  unlikely to be considered in classifications of earlier novae, and
  \citet{mor05} took the dip seen in the lightcurve
  on 2003 September 23--30 ($\Delta V\simeq 0.4-0.6$) into account
  in deriving $t_{2}$; however, this dip seems likely to be a local oscillation
  rather than a general brightness variation, given the
  global tendency of the lightcurve. The oscillation and
  deep decline of the lightcurve are typically seen in
  a subclass of moderately fast novae \citep{pay57}.}
  (\citealt{cho05}; $t_{n}$ is the time required for the nova
  to diminish by $n$ magnitudes below maximum light), which suggests
  that the speed class of V475 Sct is `moderately fast' 
  ($t_{2}=26$--$80$ d; \citealt{pay57}).
  An optical spectrum obtained around maximum brightness
  \citep{siv03,mor05} displayed many emission lines such as \ion{Fe}{2}
  and Balmer series lines, characteristic of \ion{Fe}{2} novae \citep{wil92}.
  \citet{cho05} noted the photometric and spectroscopic 
  similarities to V705 Cas and suggested that V475 Sct is a
  dusty nova \citep{geh88,mas98}.
  In this paper, we present the results of our spectropolarimetry
  of V475 Sct in its outburst phase and discuss a high-velocity component 
  seen in the polarized spectrum.

\section{Observations and Data Reduction}

\subsection{Low-dispersion spectropolarimetry}

  We took low-resolution spectropolarimetry measurements of V475 Sct 
  with HBS (the acronym stands for `Henkou-Bunkou-Sokkou-Ki,'
  which means `spectro-photo-polarimeter' in Japanese;
  \citealt{kaw99}) attached to the Cassegrain focus of
  the 0.91-m telescope at the Okayama Astrophysical
  Observatory of NAOJ on 2003 September 26, 28, and 29 UT 
  (25, 27, and 28 d, respectively).
  HBS uses a rotatable superachromatic half-wave plate and
  a fixed quartz Wollaston prism (manufactured by Oyokoden Lab.).
  Each polarization measurement consisted of several sets of four
  200-250-s exposures corresponding to the four position
  angles of the wave plate, $0\fdg 0$, $22\fdg 5$,
  $45\fdg 0$ and $67\fdg 5$.
  At the focal plane, we put two slitlets of the same
  dimension ($3\farcs 4\times 24''$). One was for the target star
  and the other was for a nearby sky region.
  Thus, we simultaneously obtained four spectra, ordinary and
  extraordinary rays of both the target and the sky,
  with a back-illuminated SITe CCD (512$\times$512, 24\micron /pixel).
  This configuration resulted in a spectral resolution
  of $\Delta\lambda\simeq 70$ \AA\ and 
  $R=\lambda / \Delta\lambda\simeq 90$.
  A log of the observations is given in Table \ref{tbl-1}.

  The data were reduced using the {\it HBSRED} package \citep{kaw99}.
  Both the instrumental polarization and depolarization
  were derived from unpolarized standard star (UP;
  \citealt{wol96,ser74}) data obtained
  with and without inserting a Glan-Taylor prism.
  The zero point of the position angle was determined from the
  observations of strongly polarized standard stars (SP;
  \citealt{wol96,sch92}) in the synthesized $V$ band.
  All the adopted calibration data were obtained during the
  2003 July--September run with the same focal slitlets as used for
  the nova observations.
  The magnitude of the derived instrumental polarization was
  approximated by a smooth function of wavelength. To give some
  examples, its magnitude was $0.363\pm 0.002$ \% at 5000 \AA\ and
  $0.393\pm 0.010$ \% at 8000 \AA\ (the error is the mean error of a
  total 31 UP observations).
  This instrumental polarization was vectorially subtracted
  from the observational data.
  The polarimetric stability, estimated from the peak-to-peak
  variation of UP polarization in the observation period
  (a total of four observations for 107~Psc and $\kappa$ Cet),
  was less than 0.1 \% over the observed wavelength range.
  A correction was also made for the zero point of the 
  position angle; its accuracy is estimated to be $0\fdg 40$,
  which is the $1\sigma$ range for a total of 11 SP observations 
  during the July--September run.

\subsection{Medium-dispersion spectropolarimetry}

  Our other spectropolarimetric observations of V475 Sct were carried out
  with the Faint Object Camera And Spectrograph
  (FOCAS; \citealt{kas02}), attached to the Cassegrain
  focus of the 8.2-m Subaru telescope \citep{kai00,iye04}
  on 2003 October 7.29 UT (36 d).
  The polarimetric module consists of a rotating
  superachromatic half-wave plate and a fixed quartz
  Wollaston prism (manufactured by B. Halle Nachfl.),
  and both the ordinary and the extraordinary 
  rays are simultaneously recorded on two MIT/LL CCDs 
  (2k$\times$4k, 15\micron /pixel ).
  We used a $0\farcs 4$ width $\times\ 20\farcs 6$ length slit
  and a 665 lines mm$^{-1}$ volume-phase holographic (VPH)
  grism \citep{ebi03},
  which gives a wavelength coverage of 5300--7700 \AA\ and
  a spectral resolution of $\simeq 2.6$ \AA\ and $R\simeq 2500$.
  The corresponding reciprocal spectral dispersion was
  $0.605$ \AA /pixel.
  To our knowledge, this is the highest resolution
  spectropolarimetry for a classical nova in the literature.
  To eliminate the contamination due to second-order spectra,
  a sharp-cut filter Y47 was used together with the VPH grism.
  The observation consisted of two sets of four sequential
  integrations at the $0\fdg 0$, $45\fdg 0$, $22\fdg 5$ and
  $67\fdg 5$ positions of the half-wave plate (Table \ref{tbl-1}).

  The frames were reduced using the standard methods for
  CCD spectroscopy.
  Since no nearby object was seen in the images we obtained, 
  we estimated the sky background component by a simple
  interpolation of the spectra at both sides of the object
  along the slit.
  The measurements of flatfield lamps through fully polarizing
  filters indicated that the depolarization factor was negligible
  ($\lesssim 0.05$) and we made no correction for it.
  The zero point of the position angle on the sky was determined 
  from the SP observation \citep{sch92}.
  The typical accuracy of the position angle is estimated to be
  $0\fdg 16$, which is the $1\sigma$ value for a total of 13 observations of
  HD~204827 in 2002 June through 2003 June.

\begin{figure}
\plotone{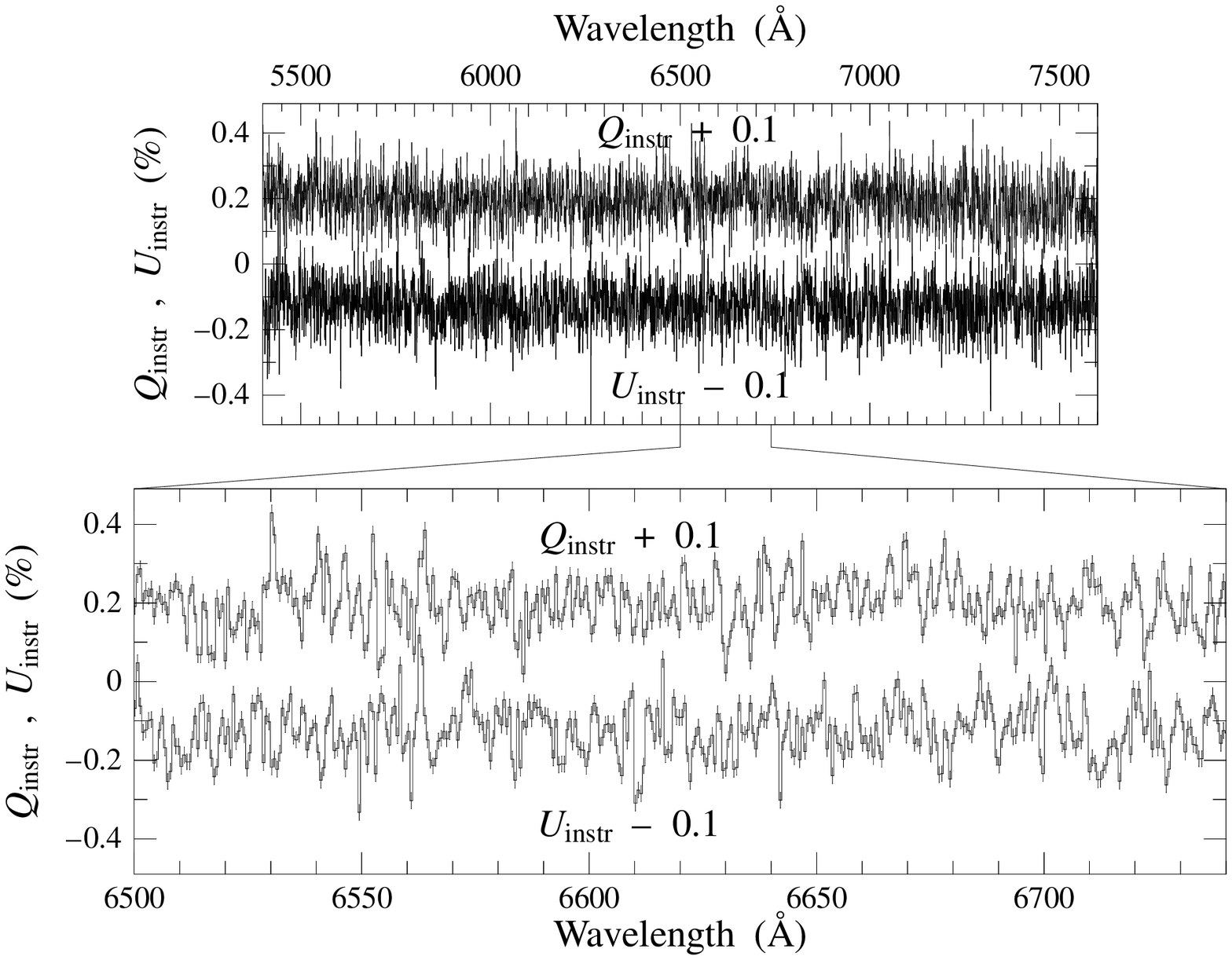}
\caption{Instrumental polarization of Subaru/FOCAS.
The data were obtained for an unpolarized standard star
HD 212311, under the same optical configuration as used for the nova.
In the observed polarization spectrum we do not see any 
significant ($>0.2$ \%) ripple pattern, which has
been a problem for some intermediate-dispersion 
spectropolarimetric observations.}
\label{fig1}
\end{figure}

  Higher-resolution spectropolarimetry ($R\gtrsim 2000$) sometimes
  suffers from a strong spectral modulation in polarization
  ({\it ripple} pattern) due to multiple reflections within the
  wave plate (e.g., \citealt{ait01,ike03,cla05}).
  In that case some method must be applied, as part of the data 
  reduction procedure, to remove the pattern (e.g., \citealt{har96}).
  Figure \ref{fig1} shows the polarization spectra obtained for
  the unpolarized star HD~212311 with the same optical
  configuration as the target observation.
  One can observe that the peak amplitude is less than $0.2$ \%\ in the
  instrumental $Q,\ U$ spectra and that the ripple pattern
  is not significant in our data.
  One possible reason for this is that the somewhat poorer surface accuracy
  of the wave plate, due to the four block construction for a
  large aperture (125 mm $\phi$), effectively reduces the
  ripple effect \citep{ait01}.
  Therefore, we were able to analyze the data in a standard way,
  except for the application of a slight correction for 
  the nearly wavelength-independent instrumental polarization
  ($Q_{\rm instr},\ U_{\rm instr}\lesssim 0.1$ \%; Figure \ref{fig1}).
  Many calibration observations during 2002--2003 suggest
  that the polarimetric stability and accuracy of Subaru/FOCAS
  are equal to or less than $0.05$ \% \citep{kaw05}.

\section{Results and Discussion}

\subsection{Spectroscopic evolution}

Figure \ref{fig2}a shows the low-dispersion
spectra of V475 Sct on 25, 27 and 28 d.
We see many prominent emission lines,
e.g., Balmer series, \ion{Na}{1}~D
$\lambda\lambda$5890,5896, \ion{O}{1} $\lambda$7773,
\ion{O}{1} $\lambda$8446, \ion{Ca}{2} IR triplet
and \ion{Fe}{2} multiplets.
At this epoch the [\ion{O}{1}] $\lambda\lambda$6300,6364
forbidden emission lines are still weak.
In the medium-dispersion spectrum obtained on 36 d
(Figure \ref{fig3}a), we can see resolved
features of these lines at 5300--7700\AA .
The \ion{Na}{1} D doublet shows complicated features
consisting of multiple P Cygni profiles, on which
sharp interstellar absorption lines are superimposed
at $\sim +75$ km s$^{-1}$.
The permitted lines,
H$\alpha$ and \ion{Fe}{2} $\lambda$6248, have P Cygni
absorption components at $-650$ and
$-1500$ km s$^{-1}$  (Figure \ref{fig4}).
The former component is likely to correspond to the 
absorption features seen at $-500$ km s$^{-1}$ on $-1$ d \citep{boe03}, 
$-480$ km s$^{-1}$ on 14 d, and $-640$ km s$^{-1}$ on 24 d \citep{cho05}.
Thus, it appears that this component was accelerated principally at 14--24 d
during the $\sim$40 post-maximum days.
However, the latter, faster component seems to
correspond to the feature that was seen at $-1140$ km s$^{-1}$ on 14 d,
and $-1370$ km s$^{-1}$ on 24 d \citep{cho05}, which implies a
monotonic acceleration for this component.
The forbidden [\ion{O}{1}] $\lambda\lambda$6300,6364 lines
are well developed by the epoch of our medium-dispersion observations.
The profile of [\ion{O}{1}] $\lambda$6300 has a flat top and
steep edges at both sides, suggesting that the lines
originate in the expanding shell (e.g., \citealt{gil99}).
The FWHM/2 of this emission line is 
$\sim 550$ km s$^{-1}$, which is comparable with the blueshift
of the P-Cyg absorption component seen in H$\alpha$ and
\ion{Fe}{2} $\lambda$6248.
These are consistent with a picture in which V475 Sct at that
epoch had a main shell expanding at a velocity of
$v_{\rm exp}\simeq 600\pm 50$ km s$^{-1}$ and a nova wind
with a speed $\sim 2.5$ times faster than the shell along the line of sight.

We note that the H$\alpha$ emission line has a considerably
asymmetric profile, with a red wing extending beyond 
$+2000$ km s$^{-1}$.
We will discuss the origin of the broad red wing in \S 3.3.

\subsection{Interstellar polarization}

Aligned nonspherical dust grains in interstellar media
produce interstellar polarization (ISP).
In order to evaluate the component of polarization intrinsic to
the nova, the ISP must be properly subtracted from the
observed polarization.
Taking into account that the position angle (PA) of
the intrinsic polarization, $\theta{\rm *}$, and that of the ISP,
$\theta_{\rm isp}$, are independent of each other, and also the fact that 
$\theta_{\rm isp}$ is almost constant as a function of wavelength, the observed
Stokes parameters $(q_{\rm obs},\ u_{\rm obs})$ may be written as
\begin{equation}
    \begin{array}{rcl}
     q_{\rm obs}(\lambda) & = &
      p_{\rm *}(\lambda )\cos 2\theta_{\rm *}(\lambda ) +
      p_{\rm isp}(\lambda )\cos 2\theta_{\rm isp} \mbox{ ,} \\
     u_{\rm obs}(\lambda) & = &
      p_{\rm *}(\lambda )\sin 2\theta_{\rm *}(\lambda ) +
      p_{\rm isp}(\lambda )\sin 2\theta_{\rm isp} \mbox{ ,}
    \end{array}
    \label{eqn1}
\end{equation}
where $\lambda$ is the wavelength, and $p_{\rm *}(\lambda)$ and
$p_{\rm isp}(\lambda)$ are the magnitudes of polarization of the 
component intrinsic
to the nova and the ISP component, respectively.
$p_{\rm isp}(\lambda)$ is found to be well represented by an empirical 
function of wavelength:
\begin{equation}
p_{\rm isp}(\lambda ) = p_{\rm max} 
\exp [ -1.15 \ln^{2}(\lambda_{\rm max}/\lambda ) ] \mbox{ ,}
\label{eqn_serkowski}
\end{equation}
where the maximum ISP, $p_{\rm max}$, occurs at the
wavelength $\lambda_{\rm max}$ \citep{ser75}.

\begin{figure}
\plotone{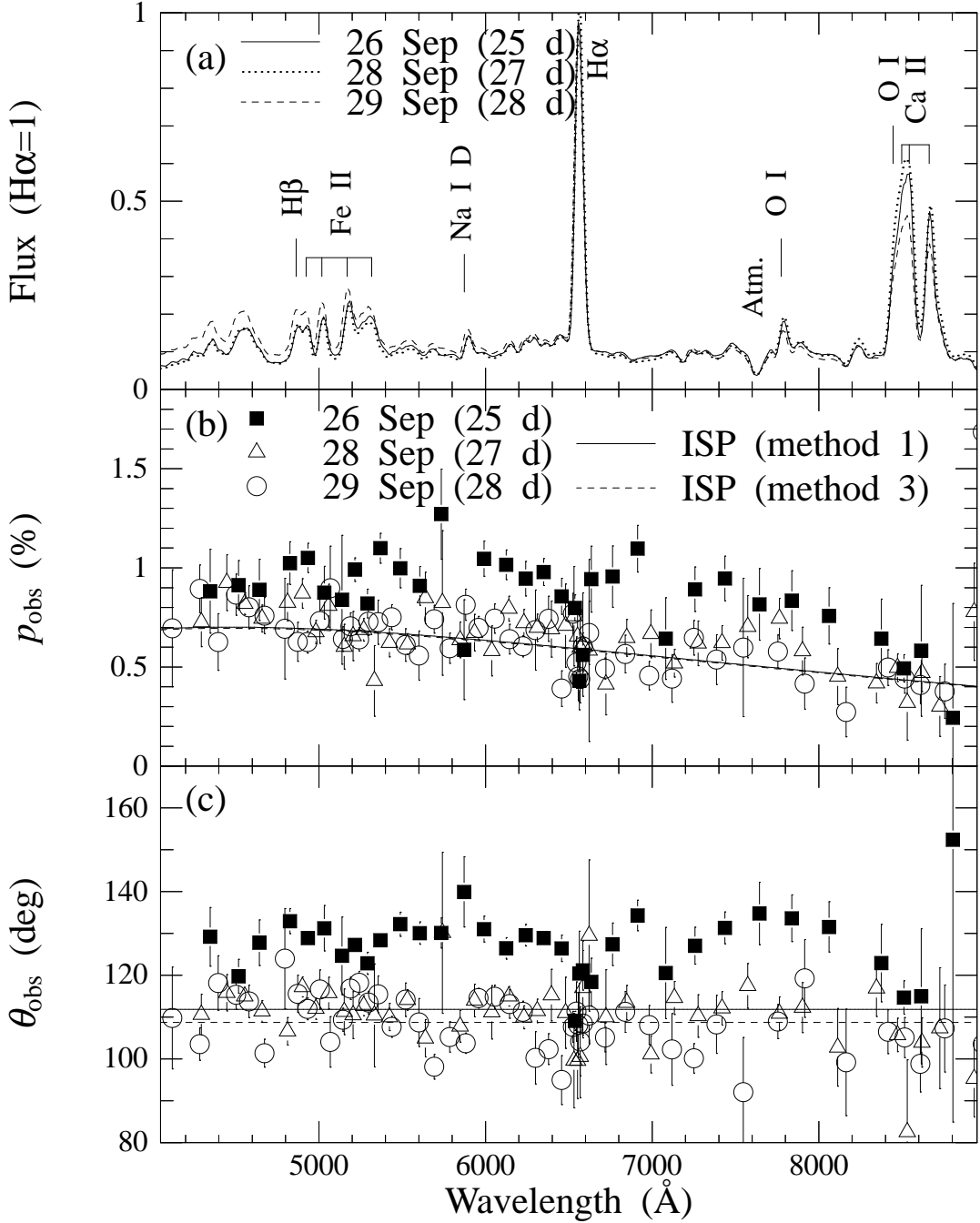}
\caption{Spectropolarimetry of V475 Sct with HBS ($\Delta\lambda=70$ \AA )
 on 2003 September 26--29 UT (25--28 d).
 From top to bottom we plot (a) the unbinned total flux, 
(b) the magnitude of the observed
 polarization, and (c) the position angle.
 Atmospheric extinction has not been corrected
 for, but the instrumental response has been calibrated.
 The major emission lines are identified by tick marks in (a).
 The polarimetric data are binned to a constant photon noise of
 $0.04$ \%, and the observational error ($1\sigma$ variation among
 ($Q$, $U$) derived for each set of four exposures) 
 in each bin is shown. The smooth solid and dashed lines in 
 (b) and (c) panels are the ISP components, estimated as described
 in the text.}
\label{fig2}
\end{figure}

\begin{figure}
\plotone{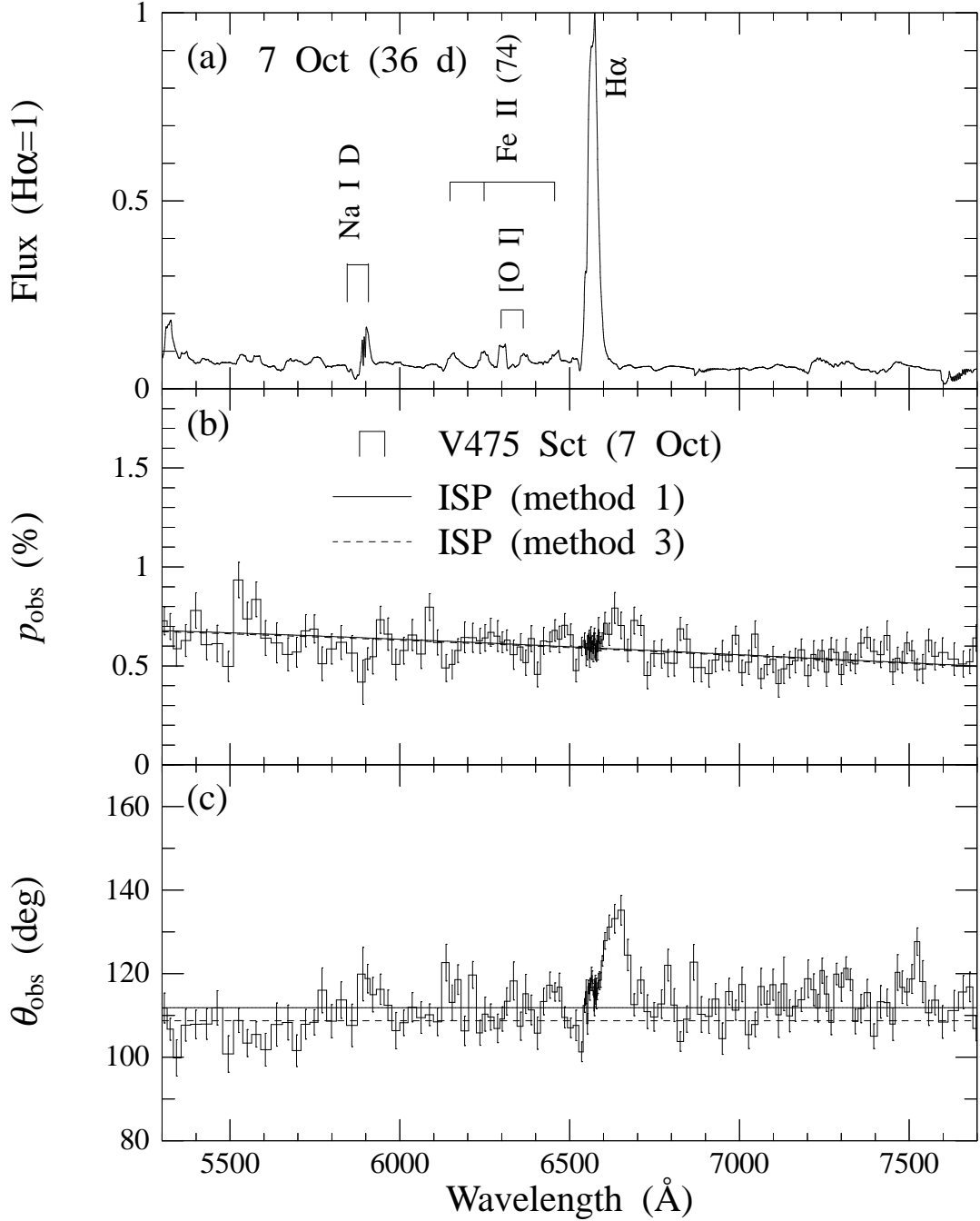}
\caption{Spectropolarimetry of V475 Sct with FOCAS 
($\Delta\lambda=2.6$ \AA ) on 2003 October 7 UT (36 d).
The data are plotted in the same manner as in Figure 2,
except that atmospheric extinction has been corrected for in (a).
The error bar denotes the observational error ($1\sigma$
variation of $Q$ and $U$ from pixel to pixel in each bin).
The bump of $\theta_{\rm obs}$ around 6640 \AA\ suggests
the existence of a distinct polarization component, while the 
small dip at 6530 \AA\ as well as the fluctuation around 6800 
\AA\ are likely to be noise-related artifacts.}
\label{fig3}
\end{figure}

\begin{figure}
\plotone{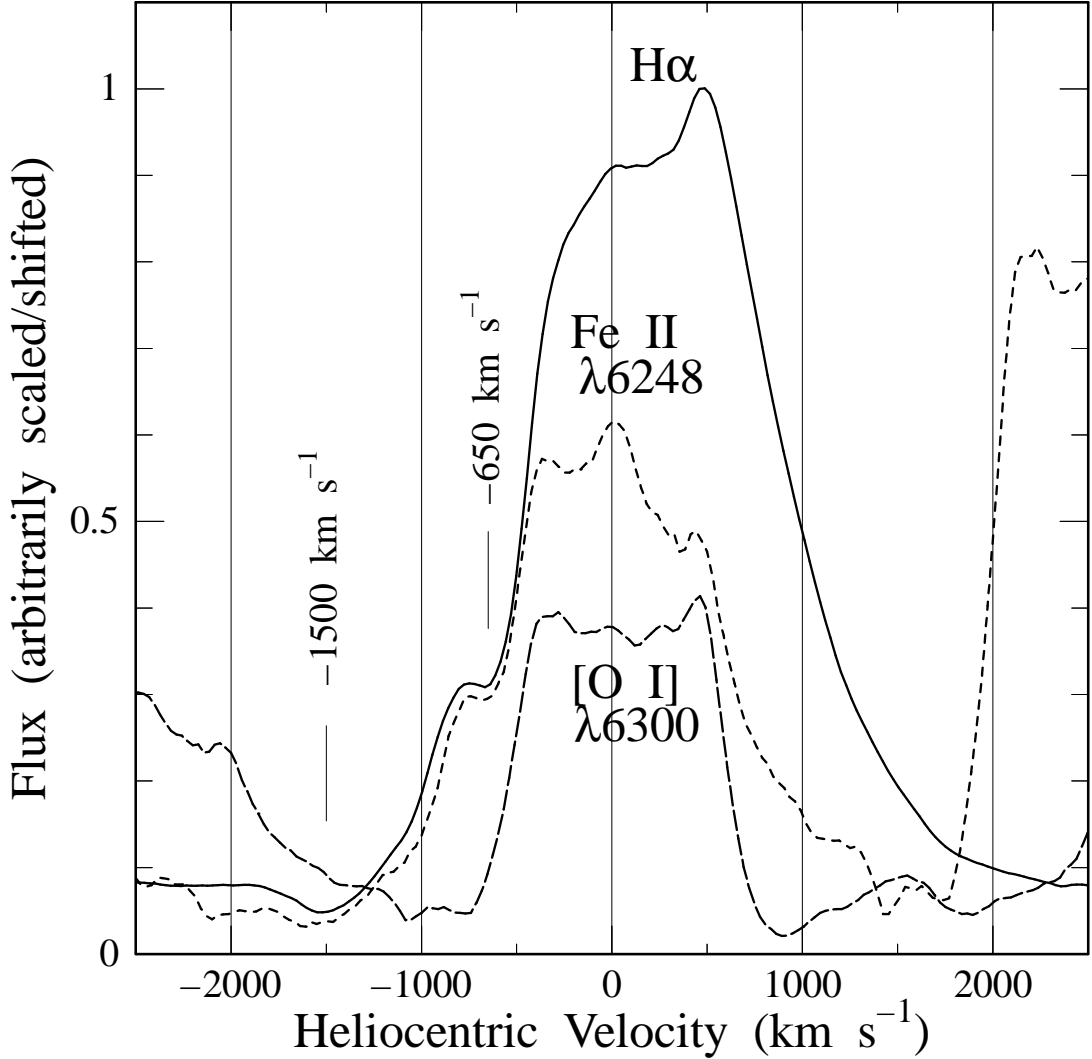}
\caption{Spectral profiles of three prominent emission lines on
36 d. The abscissa is radial velocity simply converted
from the observed wavelength. The heliocentric motion of the earth
($-29.3$ km s$^{-1}$) has been corrected.
The S/N ratio in each pixel (28 km s$^{-1}$ width) is 
80--110 with respect to the neighboring continuum light for each emission line.
The velocity resolution, estimated from the FWHM of sky emission lines,
is about 120 km s$^{-1}$ ($\simeq 2500$), suggesting
that the small bumps seen in the emission line profiles are
mostly real features.
The line widths (FWHM) are 1450 km s$^{-1}$, 1300 km s$^{-1}$
and 1100 km s$^{-1}$ for H$\alpha$, \ion{Fe}{2} $\lambda$6248
and [\ion{O}{1}] $\lambda$6300, respectively.
It should be noted that H$\alpha$ has a distinct red wing.}
\label{fig4}
\end{figure}

Figures \ref{fig2} and \ref{fig3} demonstrate the
observed polarization spectra.
A distinct polarization displacement occurred between 25 d and 27 d;
the magnitude of polarization decreased by $0.2$--$0.4$ \% and
the PA rotated by $10\arcdeg$--$25\arcdeg$ in the continuum light
at 4500--8500 \AA .
A clear depolarization effect across the H$\alpha$ emission
line was also seen on 25 d.
NLTE line emission is generally formed in the outer ionized region
and is subject to smaller Thomson scattering optical depth
compared to the continuum light from the embedded central star.
This leads to a lower polarization of line emission than
that of neighboring continuum light \citep{har68,sch92b}.
Therefore, the depolarization effect provides us with evidence that the
continuum light on 25 d was intrinsically polarized.
However, the continuum polarization remained almost
constant from 27 d through 36 d. In this period the polarization
spectra could be represented by a smooth curve, gradually decreasing
with wavelength, and the PA was almost constant
around $110\arcdeg$. There was no clear line polarization
effect across the H$\alpha$ line on 27 and 28 d.
These results suggest that the steady continuum polarization
component predominantly represents the ISP.

Here we adopt three methods to estimate the ISP toward
V475 Sct. One is to derive the almost invariable continuum polarization
on 27--36 d (method 1), the second is to average the observed
polarization in the H$\alpha$ line band on 25--28 d (method2),
and the third is to subtract the continuum component from the
result of method2 and to derive the polarization component
of the pure H$\alpha$ line emission (method 3).

For method 1, we obtained the following ISP parameters by a nonlinear
regression of Equation (\ref{eqn_serkowski}) for the polarimetric
data on 27--36 d: 
$p_{\rm max} = 0.701\pm 0.016$ \%,
$\lambda_{\rm max} = 4470\pm 120$ \AA\ and
$\theta_{\rm isp} = 111\fdg 9\pm 0\fdg 3$.
In this calculation, we neglect the H$\alpha$ line region (see method 2)
because the possibility of contamination due to
the intrinsic polarization component (as seen on 36 d) cannot be ruled out.

For method 2, we adopted the wavelength region from 6490
to 6660 \AA\ as the H$\alpha$ line band and vectorially averaged
the observed polarization. 
In this method, although the derived polarization is 
still contaminated by the continuum component, we can neglect it
because the emission line flux is much higher than the
continuum flux (e.g., \citealt{tay91,sch93,sch94}).
The derived daily values at H$\alpha$ on three nights are shown
in the upper part of Table \ref{tbl-2}, along with their
$\sigma$-weighted-mean values.

For method 3, we adopted the same wavelength region for
H$\alpha$ line band as in method 2.
In subtracting the continuum component from the line band data,
we used two wavelength regions on both sides of the line,
each of 200 \AA\ width, in which no significant polarization
feature is seen.
The derived line emission components are shown in the
lower part of Table \ref{tbl-2}.

We obtained $p_{\rm max} = 0.698\pm 0.043$ \% and 
$p_{\rm max} = 0.695\pm 0.056$ \% using method 2 and 3,
respectively, assuming that $\lambda_{\rm max}$ takes the same 
value as that of method 1.
The derived ISP parameters are summarized in Table \ref{tbl-3}.
There is no significant difference between the results derived using
the three different methods, and we will refer to the results of method 1 and 3
in the following discussion. (Method 2 gives values that are intermediate
between those of methods 1 and 3.)
The derived ISP components are plotted in Figures \ref{fig2}
and \ref{fig3}. They show that both ISP curves well
trace the continuum polarization on 27--36 d and also 
the H$\alpha$ line polarization on 25--28 d.
In Figure \ref{fig2} the magnitude of polarization at a 
data point close to the H$\alpha$ line core on 25 d is somewhat 
($1.1\sigma$) lower than the ISP curves, but we consider the 
difference at this bin to be insignificant because the width
of the bin, $\Delta\lambda=22$ \AA , is considerably narrower
than the wavelength resolution (see also Table \ref{tbl-2}).

\begin{figure}
\plotone{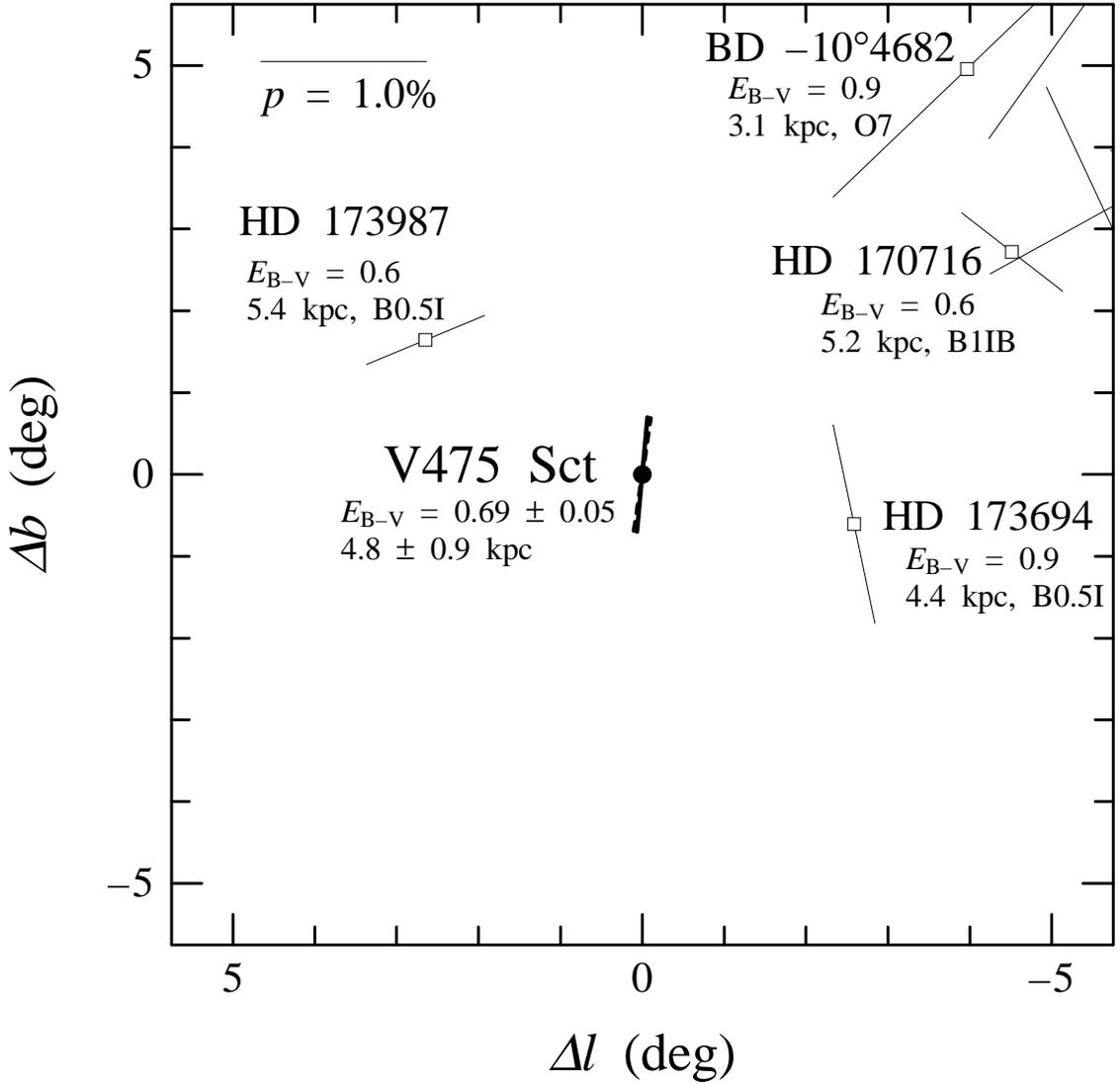}
\caption{Field star polarization map around V475 Sct.
Stars at distances of 3--7 kpc are plotted along with 
star name, color excess, distance and spectral type \citep{hei00}.
The magnitude of polarization is proportional to the length of
the vectorial bar. The horizontal bar at the top left represents
1.0 \% polarization. The ISP components estimated toward V475 Sct
are plotted at the origin by solid lines (method 1) and dashed lines
(method 3); see the text for details. 
The distance and color excess of V475 Sct are 
taken from \citet{cho05}. 
Although the polarization vectors of the field stars have a certain 
amount of scatter,
the derived ISP components seem roughly consistent with 
the nearest two stars.}
\label{fig5}
\end{figure}

Figure \ref{fig5} shows a map of the polarization of the 
field stars around
V475 Sct \citep{hei00}.
Since the distance toward V475 Sct has been estimated to be
$4.8\pm 0.9$ kpc \citep{cho05}, we selected field stars 
at 3--7 kpc.
Generally we can use the field star technique only when
we know the distance toward the target star with a
substantial accuracy and also have either a field star
containing no intrinsic polarization or plenty of field
stars to cancel out their intrinsic components by averaging.
In the present case, the error of the distance estimation 
for V475 Sct is somewhat large and we have only a few
sample stars close to the target in 3D space.
We thus use the map only for a rough check.
The three closest stars (HD 173987, HD 170716, and HD 173694)
are unlikely to have substantial intrinsic polarization because
they do not have emission lines (although they are all
supergiants and may exhibit small variability).
Although V475 Sct shows an unusually small ratio 
$p_{\rm max}$(\%)$/E_{\rm B-V}\simeq 1.0$ compared with
typical values for the Galactic interstellar medium 
($p_{\rm max}/E_{\rm B-V}\simeq 2$--$9$; Figure 9 of \citealt{ser75}), 
its ratio is nearly consistent with those of the nearest three stars,
$p/E_{\rm B-V}\simeq 1.3$--$1.4$.
This is likely to be because random components are predominant for
the Galactic magnetic fields projected to the plane of sky
in this direction, $l\sim 24\arcdeg$ (related with
the direction of the Cygnus-Orion spiral arm; \citealt{fos02})
and the polarizing efficiency against the color excess is
remarkably low.
However, when we choose two stars (HD 173987 and
HD 173694) from the viewpoint of proximity in both angular and 
radial distances, the ISP component of V475 Sct seems roughly
consistent with the ISP pattern inferred from those of the two stars.
We therefore adopted the derived ISP parameters, and calculated the
polarization component intrinsic to the nova using Equation (\ref{eqn1}).

\subsection{The red wing of H$\alpha$ emission in the polarized spectrum}

\begin{figure}
\plotone{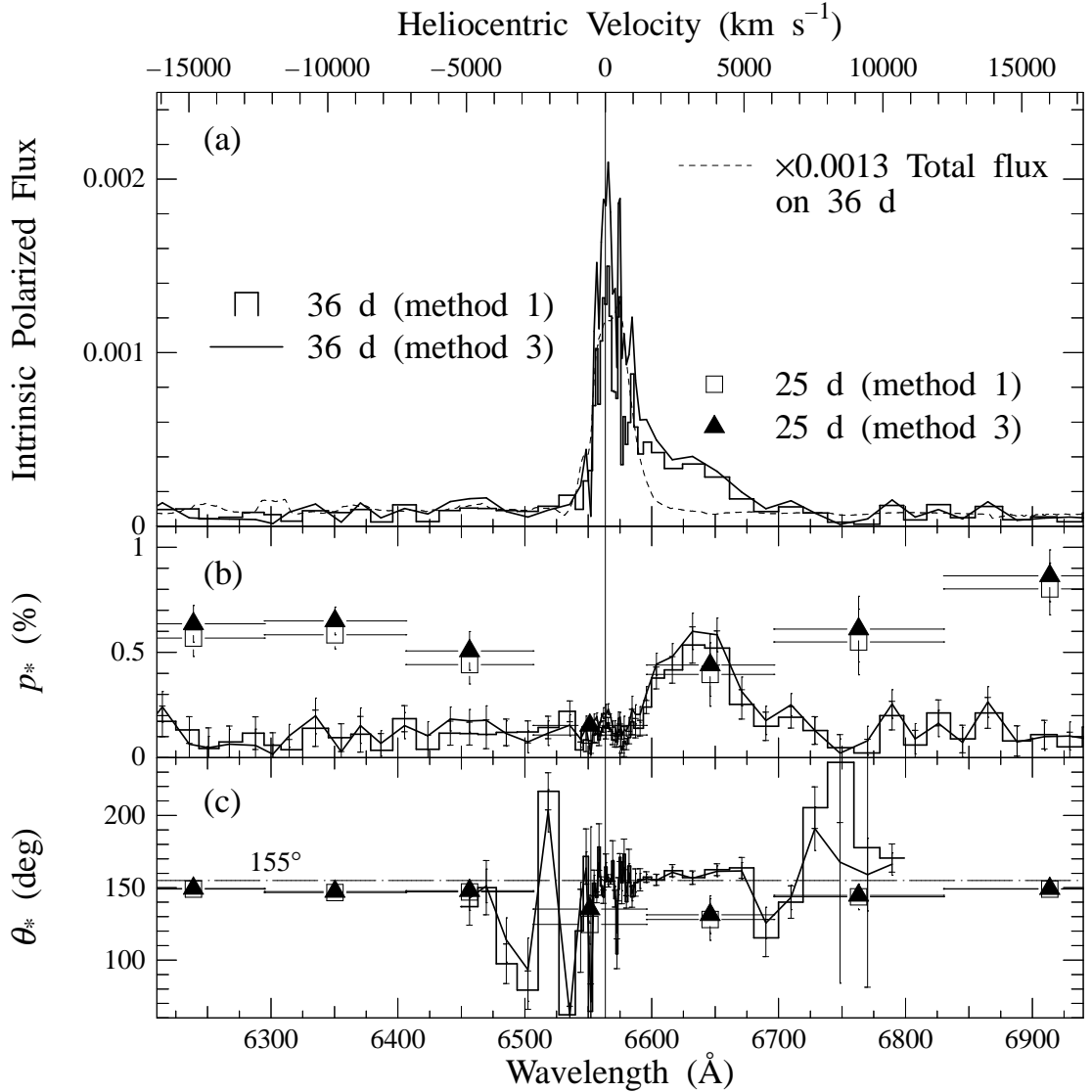}
\caption{Polarization components intrinsic to V475 Sct around H$\alpha$
emission line.
We plot (a) polarized flux (= $p_{\rm *}\times$total flux) on 36 d,
(b) the magnitude of the intrinsic polarization and (c) its PA
on both 25 and 36 d.
The data are binned in the same manner as in Figures 2 and 3,
and the vertical error bars do not include any additional
uncertainties from the ISP estimation.
In panel (a) the down-scaled total flux is plotted 
for comparison as a dashed line. 
In panels (b) and (c) the horizontal error bar
of each data point on 25 d indicates the width of the bin.
Three bins near the line core seen in Figure 2 have been
combined into one to increase the significance.
In panel (c) the data on 36 d at wavelength shorter than
6450 \AA\ and longer than 6750 \AA\ are omitted
because of lower signal to noise. 
The polarized flux has an obvious broad red wing component
at $\theta_{\rm *}\simeq 155\arcdeg\pm 15\arcdeg$; the
bump in the $p_{\rm *}$ spectrum peaks around +3500 km s$^{-1}$
and extends up to +4900 km s$^{-1}$,
suggesting the existence of a high-velocity asymmetric wind.
The intrinsic polarization of the continuum emission on 25 d 
has a well fixed PA at $\theta_{\rm *}\sim 150\arcdeg$, which is
similar to that of the wing polarization on 36 d.}
\label{fig6}
\end{figure}

In Figure \ref{fig6}, we plot the intrinsic polarization
around the H$\alpha$ line on 25 and 36 d.
There is no significant difference between the results of methods 1--3
for estimating the ISP.
In the polarization data at 25 d we can see a `pure' depolarization
effect across the H$\alpha$ emission line, in which the PA is 
almost constant against wavelength.
The small polarization of the H$\alpha$ line emission 
on 25 d ($p_{\rm *}\simeq 0.10$--$0.15$ \%) should still be
affected by the intrinsic continuum polarization.

In contrast, the polarization spectrum taken on 36 d indicates
that the H$\alpha$ line emission is clearly polarized both at the
line core and the wing component, although the continuum light
is not strongly polarized.
Two distinct properties of the emission line can be seen in Figure \ref{fig6}:
(1) The polarized flux ($p_{\rm *}\times$ total flux) on 36 d has
a broad red wing component, which corresponds to the broad bump
around $6640$ \AA\ in the $p_{\rm *}$ spectrum, and
(2) the PA of the wing polarization is well aligned, with 
$\theta_{\rm *}\simeq 155\arcdeg\pm 15\arcdeg$,
from the center wavelength of H$\alpha$ through the red edge 
of the wing $\sim 6670$ \AA .
Since all the other lines, including \ion{Na}{1} D and 
[\ion{O}{1}], were too weak to be measured polarimetrically,
we mainly discuss the H$\alpha$ line polarization in what follows.

Is the red wing component really associated with the H$\alpha$ emission?
No other strong emission line near 6640 \AA\ is known for
typical novae at similar epochs, so that the red wing
component could not be superimposed by another emission line.
In addition, the PA of the intrinsic polarization is almost constant
throughout 
the broad bump feature, including the H$\alpha$ line core.
We therefore consider that the wing component is in fact associated
with the H$\alpha$ line emission.

\begin{figure}
\plotone{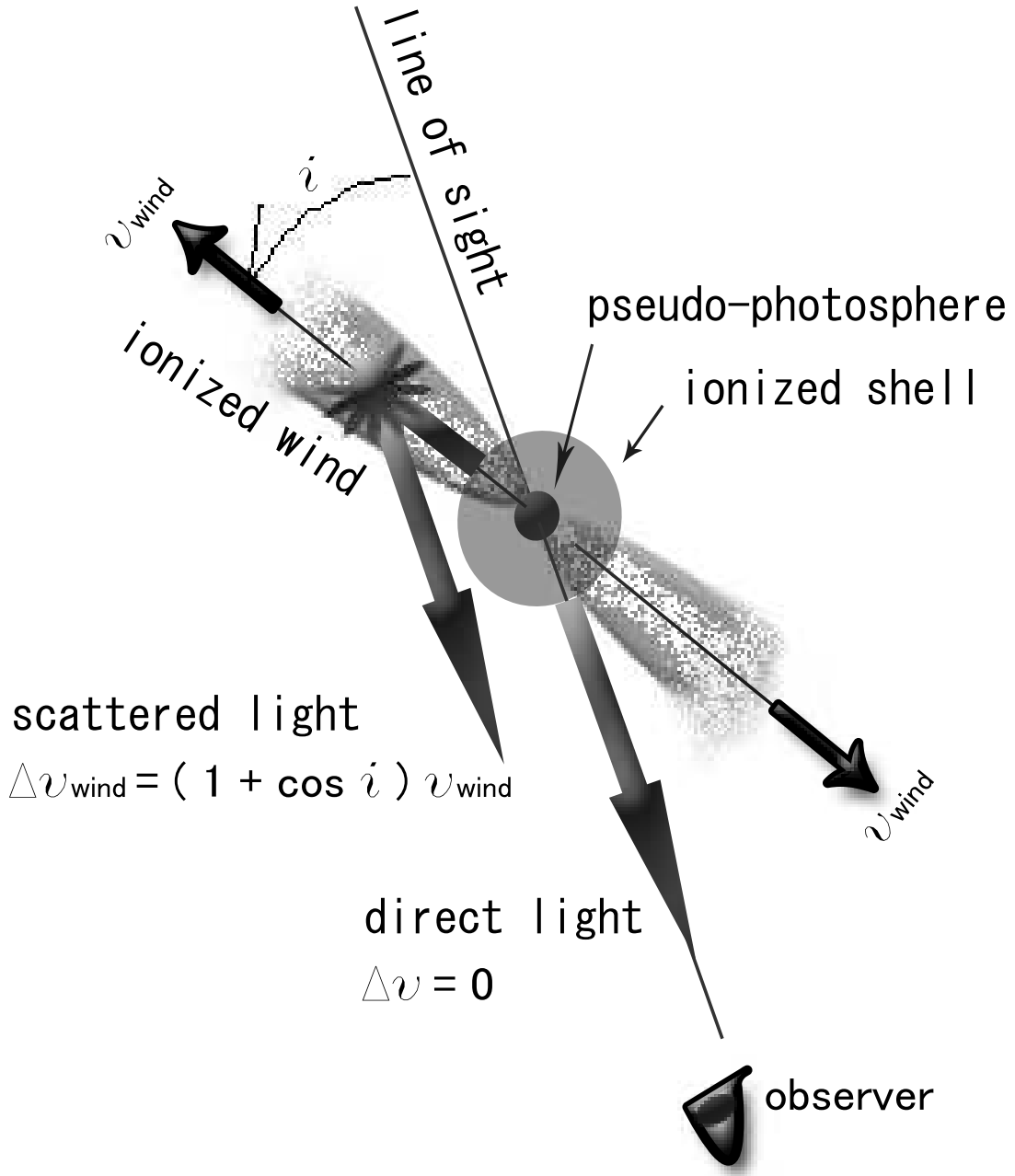}
\caption{Schematic representation of the asymmetric wind model.
The nova wind scatters the light from the central region
and produces redshifted, polarized light.}
\label{fig7}
\end{figure}

This polarized red wing component is naturally explained if
a high-velocity collimated outflow is present near the H$\alpha$ line emitting
region at PA$\simeq 65\arcdeg$ and/or $245\arcdeg$ on the projected sky.
If free electrons in a receding wind scatter a substantial amount
of light emitted from the inner (i.e., lower velocity) region,
the observed polarized flux will have a redshifted component,
owing to contributions from both scattered and unscattered
(i.e., direct) light from the inner region (see Figure \ref{fig7}).
The flux scattered by a single wind is redshifted by
\begin{equation}
 \Delta v_{\rm wind} = (1 + \cos i)\,v_{\rm wind} \mbox{ ,}
 \label{eqn_redwing}
\end{equation}
where $v_{\rm wind}$ is the velocity of the wind and $i$ is the
inclination angle of the wind direction against the line of sight
measured from the far side of the nova.
It should be noted that the flux scattered by a receding wind
is always shifted redward $\Delta v_{\rm wind} \geq 0$
and that little or no polarization is expected for $i\simeq 0\arcdeg$
or $180\arcdeg$, as the scattering angle is nearly zero.
In a realistic case, any wind is more likely to be bipolar rather
than to have a single direction of outflow.
If we assume a bipolar wind, a double-peaked profile 
(with separation $=2 v_{\rm wind} \cos i$) is expected
in the polarization spectrum.
In Figure \ref{fig6}b we find only a single broad peak
around 6640 \AA\ (corresponding to 
$\Delta v_{\rm wind}\simeq +3500$ km s$^{-1}$) that
extends up to 6670 \AA\ ($+4900$ km s$^{-1}$)
from the constancy of  $\theta_{\rm *}$.
This is consistent with a model in which the bipolar axis is 
approximately perpendicular to the line of sight
(i.e., $i\sim 90\arcdeg$).
In this case $v_{\rm wind}$ would be approximately equal to
$\Delta v_{\rm wind}$ from Equation (\ref{eqn_redwing}).
Alternatively, the single peak can be explained for any $i$ if 
we assume that one of the two scattered light components
suffers from selective extinction (e.g., more absorption
is expected for light scattered by the far-side wind component).
In this case $\Delta v_{\rm wind}$ would indicate a
lower limit on the real wind velocity, $v_{\rm wind}$.

We conclude that the H$\alpha$ wing polarization is 
more likely produced by scattering by a bipolar nova wind
with a velocity of $v_{\rm wind}\simeq +3500$ km s$^{-1}$
or higher.
To our knowledge, such a high-velocity outflow has never been
observed in moderately fast novae.
This fact does not directly indicate that V475 Sct
belongs to a peculiar category of novae.
Rather, it is plausible that many past observations lacked the
ability to detect such high-velocity outflows
because of lower spectral resolution,
insufficient time coverage, and/or lack of measurements of
polarization properties.

\subsection{Persistence of circumstellar asymmetric geometry}

Figure \ref{fig6} indicates the existence of significant
continuum polarization ($0.4$--$0.6$ \%) on 25 d.
One can see that this continuum polarization extends over
4500--8400 \AA\ with a similar magnitude and PA of polarization, 
although Figure \ref{fig6} shows only a wavelength region
near H$\alpha$.
The intrinsic continuum polarization indicates that the 
asymmetric geometry was clearly present on 25 d.
However, the polarization disappeared within the next 2 d.
Thus, the continuum polarization component was not persistent
but temporary, requiring a time-variable asymmetrical
circumstellar geometry.
We propose a plasma blob ejection as the most likely source of
the temporary polarization.
Assuming that the plasma blob ejection occurred just before
25 d, the blob would still be near the pseudo-photosphere
and would not be outside of the expanding ionized shell.
Such a blob would contribute effectively to the polarization of 
the continuum light from the pseudo-photosphere.

The similarity of the measured PA for the continuum polarization on
25 d and the wing polarization on 36 d suggests a persistence
of the asymmetric geometry during this period.
This scenario seems consistent with the polarimetric variability
found in Novae V1974 Cyg and V1494 Aql in outburst phase;
in these systems, the principal geometry of the nova wind remained
nearly constant
throughout the early decline phase, while clumpy mass ejection became
conspicuous in the later phase \citep{bjo94,kaw01}.
It remains unclear whether the plasma blob causing
the continuum polarization on 25 d also had a high velocity.
Future high-spatial-resolution imaging of the remnant of
V475 Sct may give clues to the validity of this scenario.

\subsection{Asymmetric high velocity wind}

Our observations suggest that a violent
mass-loss mechanism may exist even in a moderately fast nova.
Evidence of high-velocity outflows ($\gtrsim 2500$ km s$^{-1}$)
has been reported for some recurrent novae, U Sco, CI Aql, and RS Oph
\citep{iij02,anu00,kis01,lep99,tay89}, and also for 
ONeMg classical novae, V838 Her, V1974 Cyg, V382 Vel
and V1187 Sco \citep{ing92,sho93,sho03,cas04,lyn06}.
Those observations have shown that the peak expansion
velocity is usually recorded near maximum light and
that the velocity decreases as the optical light fades.
It is likely that in these systems the outer shell is considerably
accelerated directly after the initial ejection.
This situation seems clearly different from that
of typical classical novae, in which the expansion velocity gradually
increases after maximum light, and later spectral
systems of novae generally show higher expansion velocities
\citep{mcl60,bod89}.
It would therefore be better to discuss the wind of 
V475 Sct separately from the high-velocity outflows found in 
recurrent novae and ONeMg novae (cf. \citealt{fri93}).

Among normal classical novae (= CNO novae), only novae belonging
to the `very fast' speed class ($t_{2}<10$ d) have shown
high-velocity winds, e.g., 
$4000\pm 50$ km s$^{-1}$ in V1500 Cyg ($t_{2}\sim 2$ d; \citealt{due77}).
It is empirically known that the expansion velocity of a
classical nova becomes higher as $t_{2}$ is shorter, which has been
explained in terms of the equipartition of energy between the radiation
field and the ejected matter \citep{sha81}.
Optically thick wind calculations have shown that the
nova lightcurve and maximum wind velocity depend primarily
on the white dwarf mass, $M_{\rm WD}$, and are less sensitive to the
abundance (e.g., \citealt{kat94,kat97,kat99}), in the sense that
a more massive white dwarf usually shows a more rapid decline
in a nova outburst.
However, even for the cases with the most massive white dwarfs, 
spherically
symmetric models cannot well reproduce photospheric velocities of 
more than $\sim 2000$ km s$^{-1}$.
Calculations of nova evolutionary sequences have also given
similar results \citep{pri95}.

As mentioned in \S 1, however, evidence for 
nonspherical ejecta have been found in many novae.
\citet{kat03} applied an asymmetric jet-shaped outflow
in their steady-state mass-loss model and successfully reproduced
the high-velocity outflows with speeds up to $\sim 5000$ km s$^{-1}$
found in U Sco near maximum light.
They suggested that the terminal velocity of the outflow is 
sensitive only to the white dwarf mass, as in the spherically 
symmetric models, and that neither the composition nor the opening angle
of the jetlike outflow has a strong effect on the velocity.
The white dwarf mass of U Sco is estimated to be 
$M_{\rm WD}\sim 1.37$ M$_{\odot}$ \citep{hac00}, which is 
higher than those of moderately fast or slower class novae
($M_{\rm WD}\lesssim 1.0$ M$_{\odot}$, \citealt{liv92,pri95}).
For $M_{\rm WD}\lesssim 1.0$ M$_{\odot}$ cases,
the maximum photospheric velocity is at most 1000--1500 km s$^{-1}$
in the steady jetlike wind model \citep{kat03}.
For V475 Sct, \citet{cho05} derived $M_{\rm WD}= 0.73\pm 0.07$
 M$_{\odot}$ using their photometric data along with the
formula given by \citet{liv92}.
The existence of a steady high-velocity wind of $\sim 3500$
km s$^{-1}$ is still unusual for a moderately fast nova
without any additional acceleration source. 

We now briefly consider the possibility of a non-static
activity in a nova system within the framework of astronomical jets.
High-velocity asymmetric outflows are found in various
astronomical objects, such as T Tauri stars, X-ray binaries,
and active galactic nuclei.
\cite{liv00} reviewed past studies on jets and discussed the 
possibility that
all types of objects accompanied with jets possess accretion disks.
It has been known that some mass-accreting white dwarf systems,
such as supersoft X-ray sources and
symbiotic stars, exhibit evidence of jets (e.g., \citealt{sok04b}).
However, no convincing evidence of jets has been found so far
for classical and dwarf novae, although the configuration of
the binary system is similar to those of supersoft X-ray
sources and symbiotic stars.
\citet{sok04} applied their jet-launching thermal model with an
accretion disk to the cataclysmic binaries and suggested that
a mass-accretion rate $\dot{M}_{\rm WD}\gtrsim 10^{-6}$ M$_{\odot}$
yr$^{-1}$ is required for the formation of a shock and subsequent
mass ejection.
Their result explains why jets have not been observed
in classical and dwarf novae 
($\dot{M}_{\rm WD}\lesssim 10^{-8}$ M$_{\odot}$ yr$^{-1}$).
However, \citet{iij03} recently found a transient high
velocity component ($-2900$ and $+2830$ km s$^{-1}$) in the 
transition phase of a very fast nova V1494 Aql
104 d after the light maximum.
\citet{ret04} discussed this component together with the X-ray
activity found in the nova (e.g., \citealt{ori01,dra03}),
and suggested that the existence of jets is expected
for classical novae, owing to a possible temporary increase
of the mass-accretion rate
up to $\dot{M}_{\rm WD}\sim 10^{-6}$ M$_{\odot}$ yr$^{-1}$.
This scenario requires the existence of an accretion disk
around the white dwarf.
Although it is not well-known when the accretion disk
is reestablished after its probable destruction during
the nova outburst, some indirect evidence for the presence
of accretion disks has been found for some novae a few months or
years after the outburst \citep{lei92,ret97,ret98}.
However, for a moderately fast nova ($M_{\rm WD}\lesssim 1.0$
M$_{\odot}$) both the white dwarf and the secondary (assuming a
typical binary separation of $\sim 10^{11}$ cm) seem to be
still deeply engulfed by the photosphere $\sim 40$ d
after the outburst (Figure 12 of \citealt{kat94}).
It is not likely that the accretion of mass from the
secondary has resumed by 36 d in the case of V475 Sct.

Therefore, at the present time, we do not have any plausible explanation
for the ejection of the inferred high-velocity component.
In order to establish a realistic model, we need further 
spectropolarimetric observations and need to establish
how the continuum and line polarization 
components evolve during the outburst phase of novae.
Simultaneous radio or X-ray monitoring would be useful to
find evidence for high-energy events around the central star,
which might give rise to high-velocity ejecta.

\section{Conclusion}

We have presented low- and medium-dispersion spectropolarimetry of a
classical nova V475 Sct at four epochs in its early decline phase.
We estimated the ISP component and subtracted it from the
observed polarization.
Our main results can be summarized as follows:
\begin{enumerate}
 \item The low-dispersion spectropolarimetry revealed that the
 continuum light over most of the observed wavelength range
 was intrinsically polarized by $\simeq 0.4-0.6$ \%
 on 2003 September 26 UT (25 d).
 This polarization component almost disappeared in the next 
 2 d, requiring a time-variable asymmetric circumstellar
 geometry.
 \item The medium-dispersion spectropolarimetry on October 7 UT
 (36 d) detected an intrinsic polarization component across 
 the H$\alpha$ emission line.
 This suggests the existence of a wind with speeds of 
 $v_{\rm wind}\simeq 3500$ km s$^{-1}$ or higher. Such a fast 
 wind has never been reported in moderately fast novae.
 It is hard to explain how the wind was formed
 within the current framework of nova winds or jets.
 \item The PAs of the above two polarization components were similar to
 each other, $\theta_{\rm *}\simeq 150$--$155\arcdeg$. This suggests that
 the matter giving rise to scattering in the two cases may have 
 been ejected in a similar
 asymmetric geometry.
\end{enumerate}
Our results demonstrate the great utility of high-dispersion
spectropolarimetry for the study of nova outflows.
Further observations, however, are required.

\acknowledgments

We are grateful to staff members at Okayama Astrophysical
Observatory for their kind support.
We also thank members of the FOCAS team and staff members 
at Subaru telescope for the opportunity of the observation.
Data reduction/analysis was in part carried out on the ``sb'' computer
system operated by the Astronomical Data Analysis Center (ADAC)
and Subaru telescope of NAOJ.
This work was supported by a Grant-in-Aid from the Ministry
of Education, Culture, Sports, Science and Technology of
Japan (No. 17684004).




\clearpage

\begin{deluxetable}{ccccrc}
 \tabletypesize{\scriptsize}
 \tablecaption{Log of the observations\label{tbl-1}}
 \tablewidth{0pt}
 \tablehead{
 \colhead{UT (epoch)} & \colhead{JD} & \colhead{Telescope/} &
 \colhead{Wavelength} & \colhead{Res.} &
 \colhead{Exposure\tablenotemark{a}} \\
 \colhead{2003} & \colhead{$2,450,000+$} & \colhead{Instrument} &
 \colhead{(\AA )} & \colhead{$\lambda/\Delta\lambda$} &
 \colhead{(sec)}
 }
 \startdata
  Sep 26.53 (25 d) & 2909.03 & OAO 0.91m/HBS & 4000--9000 &   90 &
  200$\times$24 \\
  Sep 28.49 (27 d) & 2910.99 & OAO 0.91m/HBS & 4000--9000 &   90 &
  250$\times$24 \\
  Sep 29.48 (28 d) & 2911.98 & OAO 0.91m/HBS & 4000--9000 &   90 &
  250$\times$24 \\
  Oct 07.29 (36 d) & 2919.79 & Subaru/FOCAS  & 5300--7700 & 2500 &
  18$\times$4+9$\times$4 \\
\enddata
\tablenotetext{a}{The total exposure time is expressed as the integrated
 time per single frame, multiplied by the number of frames.}
\end{deluxetable}

\begin{deluxetable}{ccccc}
 \tabletypesize{\scriptsize}
 \tablecaption{Polarization component at H$\alpha$ line emission
 \label{tbl-2}}
 \tablewidth{0pt}
 \tablehead{
 \colhead{Epoch} & \colhead{Component} & \colhead{Flux (ADU)} &  
 \colhead{Polarization (\%)} & 
 \colhead{Position angle ($^{\circ}$)} 
 }
 \startdata
  25 d & Emi+Cont & $7.49\times 10^{5}$ & $0.585\pm 0.065$ & 
  $115\fdg 1\pm 3\fdg 3$ \\
  27 d & Emi+Cont & $9.68\times 10^{5}$ & $0.632\pm 0.108$ & 
  $103\fdg 8\pm 5\fdg 5$ \\
  28 d & Emi+Cont & $1.01\times 10^{6}$ & $0.581\pm 0.048$ & 
  $108\fdg 0\pm 2\fdg 4$ \\
       &          &                Mean & $0.589\pm 0.037$ & 
  $109\fdg 8\pm 1\fdg 8$ \\
  \\
  \tableline
  \\
  25 d & Emi      & $6.21\times 10^{5}$ & $0.533\pm 0.085$ & 
  $110\fdg 4\pm 4\fdg 5$ \\
  27 d & Emi      & $8.35\times 10^{5}$ & $0.650\pm 0.127$ & 
  $102\fdg 6\pm 6\fdg 4$ \\
  28 d & Emi      & $8.16\times 10^{5}$ & $0.597\pm 0.063$ & 
  $109\fdg 5\pm 3\fdg 0$ \\
       &          &                Mean & $0.587\pm 0.047$ & 
  $108\fdg 7\pm 2\fdg 3$ \\
\enddata
\end{deluxetable}

\begin{deluxetable}{cccc}
 \tabletypesize{\scriptsize}
 \tablecaption{Interstellar polarization derived using 
  three methods\tablenotemark{a}\label{tbl-3}}
 \tablewidth{0pt}
 \tablehead{
 \colhead{Method} & \colhead{$p_{\rm max}$} & 
 \colhead{$\lambda_{\rm max}$\tablenotemark{b}} &
 \colhead{$\theta_{\rm ISP}$} 
 }
 \startdata
  1 & $0.701\pm 0.016$\% & $4470\pm 120$\AA & $111\fdg 9\pm 0\fdg 3$ \\
  2 & $0.698\pm 0.043$\% & $4470\pm 120$\AA & $109\fdg 8\pm 1\fdg 8$ \\
  3 & $0.695\pm 0.056$\% & $4470\pm 120$\AA & $108\fdg 7\pm 2\fdg 3$ \\
\enddata
\tablenotetext{a}{See \S 3.2 and Equations (\ref{eqn1}) 
and (\ref{eqn_serkowski}).}
\tablenotetext{b}{Methods 2 and 3 simply use the value derived by method 1.}
\end{deluxetable}

\end{document}